# Ten Simple Rules for Reproducible Research in Jupyter Notebooks

Adam Rule[1], Amanda Birmingham[2], Cristal Zuniga[3], Ilkay Altintas[4], Shih-Cheng Huang[4,#], Rob Knight[3,5], Niema Moshiri[6], Mai H. Nguyen[4], Sara Brin Rosenthal[2], Fernando Pérez[7], Peter W. Rose[4*]

[1] Design Lab, UC San Diego, La Jolla, California, United States of America

[2] Center for Computational Biology and Bioinformatics, UC San Diego, La Jolla, California, United States of America

[3] Department of Pediatrics, UC San Diego, La Jolla, California, United States of America

[4] Data Science Hub, San Diego Supercomputer Center, UC San Diego, La Jolla, California, United States of America

[5] Departments of Bioengineering, and Computer Science and Engineering, and Center for Microbiome Innovation, UC San Diego, La Jolla, California, United States of America

[6] Bioinformatics and Systems Biology Graduate Program, UC San Diego, La Jolla, California, United States of America

[7] Department of Statistics and Berkeley Institute for Data Science, UC Berkeley, and Lawrence Berkeley National Laboratory, Berkeley, California, United States of America

[#] Current address: Biomedical Informatics Graduate Program, Stanford University, Stanford, California, United States of America

* Corresponding author

E-mail: pwrose@ucsd.edu

## Abstract

Reproducibility of computational studies is a hallmark of scientific methodology. It enables researchers to build with confidence on the methods and findings of others, reuse and extend computational pipelines, and thereby drive scientific progress. Since many experimental studies rely on computational analyses, biologists need guidance on how to set up and document reproducible data analyses or simulations.

In this paper, we address several questions about reproducibility. For example, what are the technical and non-technical barriers to reproducible computational studies? What opportunities and challenges do computational notebooks offer to overcome some of these barriers? What tools are available and how can they be used effectively?

We have developed a set of rules to serve as a guide to scientists with a specific focus on computational notebook systems, such as Jupyter Notebooks, which have become a tool of choice for many applications. Notebooks combine detailed workflows with narrative text and visualization of results. Combined with software repositories and open source licensing, notebooks are powerful tools for transparent, collaborative, reproducible, and reusable data analyses.

Introduction

The detailed and accurate descriptions of scientific methods needed to reproduce research have become increasingly difficult to provide as studies grow in scale, complexity, and reliance on computation. Numerous papers, guides, and anecdotes have highlighted the need for reproducibility in computational research and enumerated best practices [1-3], including guides in the Ten Simple Rules collection [4] and workshop materials developed by the Data Carpentry team [5]. We aim to augment this existing wellspring of advice by addressing the unique challenges and opportunities for reproducibility that arise when using computational notebooks like Jupyter Notebooks for research.

Reproducibility requires both a human- and a machine-readable description of data, software, dependencies, and the computational environment (e.g., hardware or cloud configuration) used to conduct a study along with documentation describing how all the pieces fit together. Whereas analysts previously kept this information in separate data, analysis, result, configuration, and commentary files (which were often difficult to piece together and share), they increasingly use computational notebooks such as Jupyter Notebooks and R Notebooks to combine executable code, rendered visualizations, and descriptive text in a single interactive and portable document. Jupyter Notebook in particular has seen widespread use for tracking and sharing analyses: as of September 2018, there were more than 2.8 million Jupyter Notebooks shared publicly on GitHub (https://www.github.com) [6], a number of which document academic research [7].

Jupyter Notebooks lower many barriers to reproducibility and were designed to support reproducible research by enabling scientists to craft easily shared computational narratives that mix code, results, and text [8, 9]. However, computational notebooks like Jupyter Notebook do not address all barriers to reproducibility, and introduce unique challenges of their own, many of which stem from their interactivity. An informal study [10], proposed by Mietchen, that re-ran Jupyter Notebooks mentioned in publications in PubMedCentral found only a small fraction of the sampled notebooks could be re-run without difficulty due to problems accessing data, unresolved dependencies, and platform differences.

In addition to these technical challenges, several studies have identified a lack of clear explanation in notebooks and many users' reticence to share what they feel are messy and personal artifacts [7,11]. One analysis of over 1 million Jupyter Notebooks shared publicly on GitHub found that one quarter of the notebooks had no explanatory text, and even in notebooks with text, the explanation tended to provide a high-level description of the analysis steps rather than the reasoning guiding the analysis or an interpretation of results [7]. These studies show that many scientists leverage notebooks' interactivity to create analytic playgrounds, but find it difficult to turn them into clear descriptions of research that can be read and re-run by others.

The explosive growth of computational notebooks provides a unique opportunity to support reproducible research, but the current lack of clear explanation in many notebooks and some users' resistance to sharing their messy notebooks suggests that additional tools, processes, and guidelines may be needed to achieve the vision of well-organized notebooks supporting reproducible research. Given the technical and social barriers to publishing reproducible research in Jupyter Notebooks, we have compiled a set of rules, tips, tools, and example notebooks to help guide Jupyter Notebook authors. While we focus on Jupyter Notebooks, these rules can also be applied to other documents that mix live code and narrative description, aiding in effective dissemination of results. In Fig. 1 we give a preview of the rules applied at different phases of the notebook development cycle.

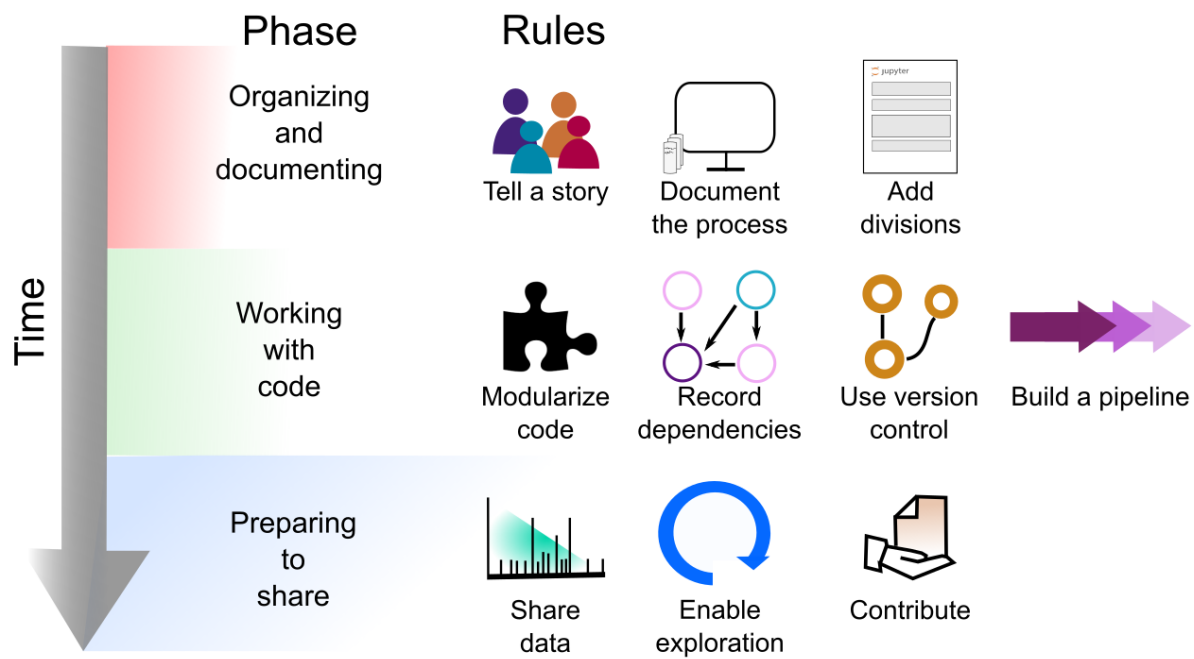

Fig. 1. Workflow for applying the ten simple rules to the creation of Jupyter Notebooks. From top to bottom we describe the three different phases needed to develop a well-documented and functional Jupyter Notebook for reproducible research. First, we organize and document the notebook (Rule 1-3). Second, the code is developed following the rules proposed here about quality standards (Rule 4-7). Finally, the notebook is made available, along with its data (Rule 8) in a manner encouraging public exploration and contribution (Rules 9-10).

## Rule 1: Tell a Story for an Audience

One key benefit of using Jupyter Notebooks is being able to interleave explanatory text with code and results to create a computational narrative [8]. Rather than only keep sporadic notes, use explanatory text to tell a compelling story that has a beginning that introduces the topic, a middle that describes your steps, and an end that interprets the results. Describe not just what you did, by *why* you did it, how the steps are connected, and what it all means.

How you tell the story will depend on your audience. Do you plan to share your notebook with a non-technical colleague in your lab, analysts at another lab, readers of a particular journal, or the general public? You may need different kinds and levels explanation for each audience. In any case, remember that your primary audience will most likely be your future self. Is your explanation clear enough that you will be able to understand and replicate the analysis *a month from now*? People often overestimate what they will be able to remember in the future, so err on the side of over-explaining. If *you* won't be able to replicate the result in the near future, how could anyone else?

## Rule 2: Document the Process, Not Just the Results

Computational notebooks' interactive nature makes it quick and easy to try out and compare different approaches or parameters – so quick and easy that we often fail to document those interactive investigations at the time we perform them. Thus, the advice long provided regarding paper lab scientific notebooks becomes even more critical: make sure to document all your explorations, even (or perhaps especially) those that led to dead ends! These will help you remember what you did and why. You can always remove them later if turning your notebook into a pipeline (See Rule 7) or preparing to share it with a different audience (Rule 1).

Many notebook users wait to add such explanatory text until the end of an analysis, after they have a solid result. Don't wait – by that point you may have forgotten why you chose a particular parameter value, where you copied a block of code from, or what you found interesting about an intermediate result. If you do not have time to fully document what you are doing or thinking in the moment, leave short descriptive notes to remind yourself what to add when you get to a good stopping point. While the code needed to reproduce the analysis is automatically captured in your notebook, the reasoning and intuition are not. It is okay if the story in your notebook changes over time; you should still tell a story from the very beginning, even if you don't know the ending yet.

Clean, organize, and annotate your notebook after each experiment or meaningful chunk of work (but be sure to track these changes in case of any mistakes – see Rule 6!). When preparing to publish, if possible avoid manually tweaking figures with desktop publishing tools and instead using plotting libraries with the notebook to produce publication-ready versions of figures and other artifacts to be used in manuscripts. Make sure you include your name as well as contact information for yourself and a future contact in your lab that can answer basic questions about the code. Documenting the beginning and end date of your analysis is also a good idea and can highlight the effort that you have put into the development of the notebook.

## Rule 3: Add Divisions to Make Steps Clear

Notebooks are an interactive environment, so it is very easy to write and run one-line cells. This supports experimentation but can leave your notebooks messy and full of short fragments that are hard to follow. Instead, try to make each cell in your notebook perform one meaningful step of the analysis that is easy to understand from the code in the cell or the surrounding markdown description. Modularize your code by cells and label the cells with markdowns above the cell. Think of each cell as being one paragraph, having one function, or accomplishing one task (e.g., create a plot). Avoid long cells (we suggest anything over 100 lines or one page is too long). Put low-level documentation in code comments. Use descriptive markdown headers to organize your notebook into sections that can be used to easily navigate the notebook, and add a table of contents. Split long notebooks into a series of notebooks and keep a top-level index notebook with links to the individual notebooks.

## Rule 4: Modularize Code

It is always good practice to avoid duplicate code, but in notebooks it is especially easy to copy a cell, tweak a few lines, paste the resulting code into a new cell or another notebook, and run it again. This form of experimentation is expedient, but makes notebooks difficult to read and nearly impossible to maintain if you want to change the functionality of or fix a bug in the copied code. Instead, wrap code you are about to copy and reuse in a function, which you can then call from as many cells as desired. If you are going to reuse the code in other projects or notebooks, consider turning it into a module, package, or library – and following good software development practices like unit testing.

Not only does modularization save space, support maintenance, and ease debugging, it also makes it easier to add interactivity. For example, you can tie widgets (ipywidgets, https://ipywidgets.readthedocs.io/en/stable/) to functions to support exploration of different parameter values or support interaction with visualizations without needing to modify the code.

## Rule 5: Record Dependencies

Reproducing your analysis in the future will require accessing not only your code, but also any dependencies. As is best practice across computational science, manage your dependencies explicitly from the start using a tool such as Conda's environment.yml or pip's requirements.txt, to list all relevant dependencies (including *their* software versions). Always conduct your work in an environment created only from these dependencies to ensure you do not add undocumented dependencies.

In notebooks, you can explicitly print out your dependencies using a notebook extension such as watermark (https://github.com/rasbt/watermark). Listing the versions of critical dependencies in the notebook itself (best done at the bottom) will ensure that, if used in isolation from its environment, the notebook still contains critical information to help readers replicate results.

## Rule 6: Use Version Control

Version control is a critical adjunct to notebook use, because the interactive nature of notebooks makes it easy to accidentally change or delete important content. Furthermore, since notebooks contain code, and code inevitably contains bugs, being able to determine the history of when a given bug you have discovered was introduced to the code vs when it was fixed – and thus what analyses it may have affected – is a key capability in scientific computation. Consult the Ten Simple Rules paper by Perez-Riverol et al. [12] on how to take advantage of Git and GitHub for version control generally. Also follow best practices for organizing your repository for easy version control, for example, http://drivendata.github.io/cookiecutter-data-science/.

However, be aware that Jupyter Notebooks store both code and specialized and extensive metadata about each cell as a text file in the JSON (JavaScript Object Notation) format. Version control systems compare differences in these JSON files, not differences in the user-friendly notebook GUI (graphical user interface). Because of this, reported differences between versions of a given notebook are usually difficult for users to find and understand, because they are expressed as changes in the abstruse JSON metadata for the notebook. One way to address this issue is to use a notebook-specific diffing tool like nbdime that understands notebook structure and presents diffs in meaningful ways (https://github.com/jupyter/nbdime). Another is to use a post-save hook to convert notebook files into a format more amenable to comprehensible versioning (https://www.svds.com/jupyter-notebook-best-practices-for-data-science/). For example, you may keep two folders, one with your raw notebooks and one with you notebooks converted to .py files stripped of output; however, remember to version control both, even if you depend on the stripped version to easily see what the versioned changes were!

## Rule 7: Build a Pipeline

Notebooks documenting initial, exploratory investigations will rarely be widely generalizable, but once a stable analysis approach has been identified, a well-designed notebook can be generalized into a pipeline that easily repeats that analysis using different input data and parameters. With this in mind, design your notebook from the beginning to allow such future repurposing. Place key variable declarations, especially those that will be changed when doing a new analysis, at the top of the notebook rather than burying them somewhere in the middle. Perform preparatory steps, like data cleaning, directly in the notebook, and avoid manual interventions when possible.

Because notebooks' interactivity make them vulnerable to accidental overwriting or deletion of critical steps by the user, if your analysis runs quickly, make a habit of regularly restarting your kernel and re-running all cells to make sure you did not accidentally delete a step while cleaning your notebook (and if you did, retrieve the code for it from version control). To allow partial execution of complex analyses, break long notebooks into smaller notebooks that focus on one or a few analysis steps. Then, ensure that each notebook stores serialized versions of key intermediate results to disk for subsequent notebooks to use.

Once a notebook has been developed, it can be parameterized with a tool such as papermill (https://github.com/nteract/papermill). Such notebooks can be used not only interactively, but also as command-line tools that can be executed automatically--a great boon for pipelines! Consider linking your analysis pipeline steps via a Makefile or similar tool that allows for complete non-interactive execution of the entire pipeline, either in full or partial steps. Such automation also supports code quality techniques like software testing; consider testing your workflows from end to end each time a change is committed by integrating your repository to a Continuous Integration system (e.g., https://travis-ci.org/). Last but not least, be aware that pipeline notebooks will almost certainly have a very different story (Rule 1) than the initial analyses that engendered them! Remember to remove any introduction, interpretation, or conclusion text that is not universally applicable to different inputs and results, and instead replace it with guidance for the pipeline user on how to run and interpret its (potentially novel) results.

## Rule 8: Share and Explain Your Data

Having access to a clearly-annotated notebook is of little use to reproducibility if the underlying data is locked away. Strive to make your data, or a sample of your data, publicly available along with the notebook. Notebooks make it easy to provide a description of your input data and upstream processing steps, which are essential for interpreting results.

Ideally, you will share your entire dataset alongside your notebooks. We realize many datasets are too large or too sensitive to share this way. In these cases, consider breaking down large and complex datasets into tiers such that, even if the raw data is prohibitively large to include alongside your published notebooks or is constrained by privacy or other access issues, reproducibility isn't lost. You can host public copies of medium-sized, anonymized data in a variety of hosting services (e.g., figshare (https://figshare.com/), zenodo (https://zenodo.org/)), and include further processed datasets alongside the notebooks in the final repository. To uniquely and permanently identify datasets, another important aspect of reproducibility, these hosting services provide Digital Object Identifiers (doi). This tiered approach both provides public confidence and allows others to replicate and reuse the latter stages of an analysis even without access to the full, raw dataset.

## Rule 9: Enable Your Notebooks to Be Read, Run, and Explored

If you have followed the previous rules, your notebooks should capture your entire process and be easy to read. But how will others access, run, and explore them? There are a number of ways you can support others' reuse of your notebooks. First, store your notebooks in a public code repository with a clear README file and a liberal open source license (https://opensource.org/licenses) granting permission to reuse your code.

Beyond granting permission to reuse, consider how you can leverage the unique structure of notebooks to support reading and exploration. At the very least, leave static HTML/PDF versions of all notebooks stored in the final version of the repository accompanying a publication. If in 20 years all other execution technology fails, these are likely to still provide a readable archival record, and with a full dependences list, future users are more likely to be able to recreate the compute environment. You can also use Nbviewer (https://nbviewer.jupyter.org/) to provide static views of your executed notebook without needing to convert it to a PDF/HTML document first. GitHub uses this service to render any notebooks on their site, so pushing a notebook to GitHub is another good way to make static views easily available.

To support others running your notebooks, consider adding widgets as mentioned in Rule 4 so they can explore your data and new parameters without writing code (ipywidgets).  You can use Binder [13] to provide a zero-install environment to run your notebooks in the cloud (https://mybinder.org/) using Jupyter Notebook or Jupyter Lab for community members who would find installation a barrier to use. More generally, you can create a Docker image of your environment (https://docs.docker.com/) to ease setup.

## Rule 10: Contribute to Reproducible and Open Research

Clearly, the mere use of a computational notebook does not guarantee reproducibility.  If the convenience and interactivity of this technology has convinced you to adopt it, take the next step and become an advocate in your lab or workplace in promoting its reproducible use. Ask lab-mates or colleagues to try to run one of your notebooks, and then listen when they explain what went wrong. Try to run their notebooks and let them know if you hit snags. Commit yourself to reproducibility as key element of all your research group's computational work, not a phase performed after an analysis is complete or an afterthought triggered by journal or reviewer demands.

Much of the infrastructure underlying reproducible research including Jupyter Notebook, many Python data science libraries, and the R tidyverse are open source. Many are free to use only because (often volunteer) contributors and maintainers dedicate time working on the project. Consider contributing to an open-source code base dedicated to supporting reproducible research or building your own software to support reproducibility in your line of research. (Prlić and Procter [14] discuss considerations, impact, and benefits of contributing to an open source project.)  Share and publish your efforts following some of the principles described in this and related Ten Simple Rules papers.

## Annotated Notebooks

To demonstrate the ten rules, we have created a Git Repository with annotated example notebooks (https://github.com/jupyter-guide/ten-rules-jupyter).  Following Rule 8, read, run, and explore these notebooks. In addition, we have created a repository (https://github.com/jupyter-guide/jupyter-guide) to crowdsource more technical and in-depth tutorials and to keep up with the rapidly evolving Jupyter ecosystem. We encourage you to contribute and share your experiences and know-how following Rule 10.

# Conclusions

Reproducibility lies at the heart of science, and several papers have already provided excellent general advice for how to perform and document computational science in ways that make it easier to reproduce. However, the advent of computational notebooks presents new opportunities and challenges for reproducibility, both easing precise documentation of complex workflows, and complicating it by means of interactivity. We present ten simple rules for reproducible research in computational notebooks, focusing on annotation of the analysis, organization of code, and ease of reuse. Informed by our experience, but we hope they contribute to the ecosystem of individuals, labs, publishers, and organizations working to make reproducible research a reality.

# Acknowledgements

This paper represents a summary of the workshop "Reproducible Research and Interactive Education - Application of Jupyter Notebooks" held at UC San Diego on April 5, 2018. We thank all participants who contributed ideas to this paper including Tiago Leao, Nathan Mih, Shweta Purawat, Michael Reich, Britton Smith, Shuai Tang, and Guorong Xu.